\documentclass[twocolumn]{aastex63}
\usepackage{soul}
\usepackage{mathrsfs}
\usepackage{multirow}

\shorttitle{Measuring IXPE Crab Polarization}
\shortauthors{Wong et al.}

\graphicspath{{./}{figures/}}
\usepackage{amsmath,amstext}

\begin{document}

\title{Improved Measurements of the IXPE Crab Polarization}

\author[0000-0001-6395-2066]{Josephine Wong}
\affiliation{Kavli Institute for Particle Astrophysics and Cosmology, Stanford University, 452 Lomita Mall, Stanford, CA 94305, USA}
\affiliation{Department of Physics, Stanford University, 382 Via Pueblo Mall,
Stanford CA 94305}
\email{joswong@stanford.edu} 

\author[0000-0001-6711-3286]{Roger W. Romani}
\affiliation{Kavli Institute for Particle Astrophysics and Cosmology, Stanford University, 452 Lomita Mall, Stanford, CA 94305, USA}
\affiliation{Department of Physics, Stanford University, 382 Via Pueblo Mall,
Stanford CA 94305}

\author[0000-0002-6401-778X]{Jack T. Dinsmore}
\affiliation{Kavli Institute for Particle Astrophysics and Cosmology, Stanford University, 452 Lomita Mall, Stanford, CA 94305, USA}
\affiliation{Department of Physics, Stanford University, 382 Via Pueblo Mall,
Stanford CA 94305}

\begin{abstract}

X-ray polarization from the Imaging X-ray Polarimetry Explorer ({\it IXPE}) provides an important new probe of the geometry of the pulsar emission zone and of particle acceleration in the surrounding pulsar wind nebula (PWN). However, with {\it IXPE}'s modest $\sim\,20-30^{\prime\prime}$ spatial resolution, separation of the pulsar signal from the nebula is a challenge. Conventional analysis defines an ``off'' phase window as pure nebular emission and subtracts its polarization to isolate the phase-varying pulsar (``on-off fitting"). We present a more sensitive scheme that uses external measurements of the nebula structure and pulsar light curve to isolate their contributions to the phase- and spatially-varying polarization via least-squares regression (``simultaneous fitting"). Tests with simulation data show $\sim$30\% improvement in pulse phase polarization uncertainties, decreased background systematics, and substantially improved nebular polarization maps. Applying ``simultaneous fitting" to early {\it IXPE} Crab data extracts additional phase bins with significant polarization. These bins show interesting departures from the well-known optical polarization sweeps, although additional exposure will be needed for precise model confrontation.

\end{abstract}

\section{Introduction} 
\label{sec:intro}

The Imaging X-ray Polarimetry Explorer ({\it IXPE}) \citep{Weisskopf:2022} is NASA's first satellite dedicated to measuring X-ray polarization. Since it was launched in Dec 2021, it has observed several pulsar wind nebula (PWN), including Crab, Vela, MSH 15-5(2), and B0540. In these sources, a central pulsar is embedded in a bright halo of relativistic charged particles accelerated by the pulsar's spin-down. The pulsed emission is generally thought to originate from the magnetically-controlled flow within and just beyond the light cylinder. Far from the light cylinder, particles and fields form a randomized wind that powers the nebula (the PWN). The nature of particle acceleration in both regions is still not fully understood although many PSR/PWN models have been proposed. For PWN reviews, see \citet{2006ARA&A..44...17G} and \citet{2020arXiv200104442A}. {\it IXPE} measurements of synchrotron X-ray polarization allows us to probe the underlying magnetic structure of these zones and test these models. 

Due to {\it IXPE}'s modest HPD (Half Power Diameter)$\sim 20-30^{\prime\prime}$ spatial resolution, isolating the pulsar signal from the surrounding nebular emission is a challenge. The conventional method is to assume an ``off" phase of pure nebular emission and subtract it to isolate the varying pulsar contribution \citep{2023NatAs.tmp...74B}. In this paper, we introduce a more sensitive method using Chandra measurements of the nebula map and pulsar light curve, with high spatial and time resolution, to weight their respective contributions to the phase- and spatially-varying polarization, extracting the polarization properties via least-squares regression. In Section 2, we describe this method and demonstrate, using simulated Crab observations, that it produces an improved nebula polarization map and more accurate pulsar polarization curve. In Section 3, we apply this technique to the {\it IXPE} Feb/Mar 2022 observation of the Crab and comment on new features visible in the data. In Section 4, we discuss notable differences between our X-ray measurements and previous optical polarization measurements and further extensions of our work. We conclude with brief remarks in Section 5.

\newpage
\section{Simulation Analysis}
\label{sec:simulation}

We used {\it IXPE}'s simulation and analysis software (\texttt{IXPEsim V14.3.1} and \texttt{IXPEobssim V28.1.0}, using V11 response functions) \citep{Baldini:2022} to generate a mock 100ks observation of the Crab PWN. In the simulation, the pulsar light curve and phase varying spectrum are derived from Chandra ({\it CXO})  HRC/LETG observations \citep{Weisskopf:2004} and the nebula image comes from a contemporaneous {\it CXO} ACIS image (OBSID 23539, PI Slane) obtained to support the {\it IXPE} study. 

ACIS CCD limitations produce two artifacts in the image. First, photons collected during readout produce trails along the charge-transfer direction: a bright narrow narrow streak from the pulsar and a diffuse rectangular patch from the nebula. We excised the streak, replacing it with a random sample of events, half each from two streak-sized regions above and below. The streak is close to the nebula symmetry axis, so this gives a fairly continuous flux and spectrum in the corrected image. The diffuse flux from the nebula was corrected by defining a nebula boundary, marking the zones extended to either side along the readout axis and replacing events in these zones with events sampled from a readout-free background region. Second, CCD pile-up eliminates counts in the pulsar core and leaving an annulus of pulsar counts in the PSF wings. We excised a circular region of radius $\sim$3.2$^{\prime\prime}$ and replaced it with photons drawn from a $\sim$18$^{\prime\prime}$-wide surrounding hemisphere (excluding flux at the base of the jet). 

The {\it IXPE} detector reads out pixels directly without charge transfer (and with small $\sim 1.1$\,ms dead time/event). Thus our corrected image better matches, in morphology and spectrum, the nebular flux that {\it IXPE} records. To this corrected PWN image, we add the phase-variable pulsar point source, with photons spatially distributed according to the {\it IXPE} PSF.

\begin{figure}
    \hspace*{-1cm}                                                     
    \includegraphics[width=1.2\linewidth]{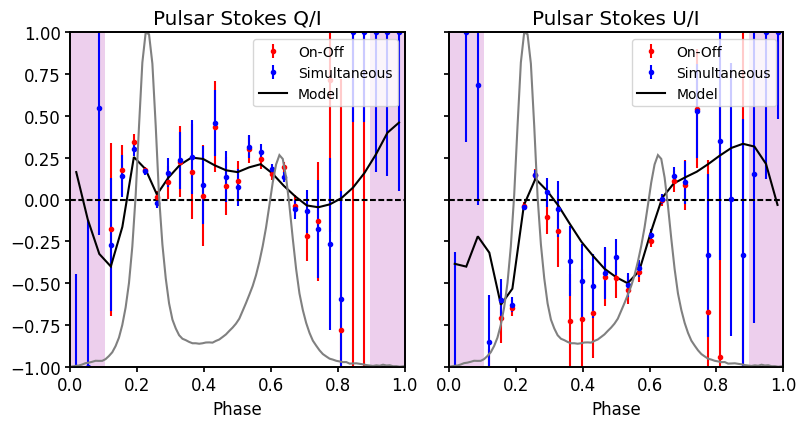}
    \caption{On-off and simultaneous fits for the pulsar polarization of a simulated 100ks Crab observation. The black line represents the polarization model. The light curve is displayed in the background (gray) with the off-pulse phase region shaded in purple. By definition, $Q_{\rm pulsar} =  U_{\rm pulsar} = 0$ in these phase bins for on-off analysis.}
    \label{fig:pulsar_sim}
\end{figure}  

To complete the simulation, we assembled a plausible model of the X-ray polarization. For the pulsar, we used OPTIMA optical polarization measurements \citep{Slowikowska:2009} but compressed the sweep so that the position relative to the optical peaks was mapped to the X-ray peaks. For the nebula, we assumed that magnetic field lines follow elongated features (e.g. toroidal near the central wisps, arced with the limb at the outskirts, and parallel to the jet) to construct a polarization map.

In the on-off method, we divided the data into 29 equal-spaced phase bins, with $0.897-1.103$ (6 phase bins, see Fig. \ref{fig:pulsar_sim}) as the off-pulse phase. For the pulsar polarization, we restricted our analysis to a circular 20$^{\prime\prime}$ radius aperture centered on the pulsar to minimize nebula contamination and a single energy bin between 2-8 keV ({\it IXPE} optimal energy range). To determine the nebula polarization, we used 5 energy bins and a 150$^{\prime\prime}$ $\times$ 150$^{\prime\prime}$ area, divided into a $15\times15$, 10$^{\prime\prime}$ pixel grid, mapping the flux from the off-pulse phases. This sub-HPD pixel grid gives a more detailed map of the polarization morphology of the nebula, although the polarization value assigned to each pixel will be slightly influenced by adjacent pixels. This may be mitigated by PSF-based deconvolution of the final maps.

In the simultaneous method, we used the latter binning scheme and solved for the desired $q$ and $u$ of the pulsar phase-resolved and nebula spatially-resolved spectra using the observed $Q$ and $U$ fluxes as:
\begin{equation}
\begin{split}
    Q_{ijklm} &= I_{{\rm psr}, ijklm} \times q_{{\rm psr}, ij} + I_{{\rm neb}, ijklm} \times q_{{\rm neb}, jkl} \\
    U_{ijklm} &= I_{{\rm psr}, ijklm} \times u_{{\rm psr}, ij} + I_{{\rm neb}, ijklm} \times u_{{\rm neb}, jkl}
\end{split}
\end{equation}

\noindent where the indices $i$, $j$, $k$, and $l$ represent the phase, energy, and spatial position of the bin and $m=1-3$ refers to the three IXPE telescopes. Assuming equally-spaced phase bins,
\begin{equation}
\begin{split}
    I_{{\rm psr}, ijklm} &= \mathscr{I}_{{\rm psr}, ijm} \times \mathrm{PSF}_{jklm} \\
    I_{{\rm neb}, ijklm} &= \mathscr{I}_{{\rm neb}, jklm}\ /\ i_{\rm max} .
\end{split}
\end{equation}

\noindent where $\mathscr{I}_{{\rm psr}, ijm}$ and $\mathscr{I}_{{\rm neb}, jklm}$ are the expected counts determined from {\it CXO} measurements of the phase-dependent pulsar spectrum and the energy-resolved image of the nebula, passed through \texttt{IXPEobssim} to account for the instrument response. The three {\it IXPE} telescopes have significantly different PSFs (and slightly different effective areas), hence the $m$ dependence. In practice, we use a long 1Ms simulation to predict the counts for a shorter observation to reduce statistical errors.

We want to find the parameters that minimize the Gaussian error, where the variance of $Q_{ijklm}$ and $U_{ijklm}$ are given by \citealt{Kislat:2015}:
\begin{equation}
\begin{split}
    \mathrm{var}(Q)  &= N(\frac{2}{\mu^2}-\bar{q}^2) \\
    \mathrm{var}(U)  &= N(\frac{2}{\mu^2}-\bar{u}^2) \\
    \mathrm{cov}(QU) &= -N\bar{q}\bar{u}
\end{split}
\end{equation}

We have now constructed an over-determined least-squares problem for which a best-fit solution must exist. The fitting was performed using the scipy function, \texttt{lsq\_linear}, which allows for the specification of parameter bounds. We found that it was sometimes helpful to introduce physical limits $\{-1, 1\}$ on the Stokes parameters to obtain a good fit to the model. Because bounded-value least squares is an iterative optimization algorithm, when values reach the physical $q,u \in \{-1,1\}$ bounds, the error bars could not be obtained analytically. To handle such cases, we recovered error bars with bootstrap analysis. We found that 500 bootstrap iterations was enough for the uncertainties to converge to within 1.5\%, and confirmed that, when bounded value least-squares converged without hitting the bounds, standard error propagation gives accurate fit errors.

\begin{figure*}
    \centering
    \includegraphics[width=.99\textwidth]{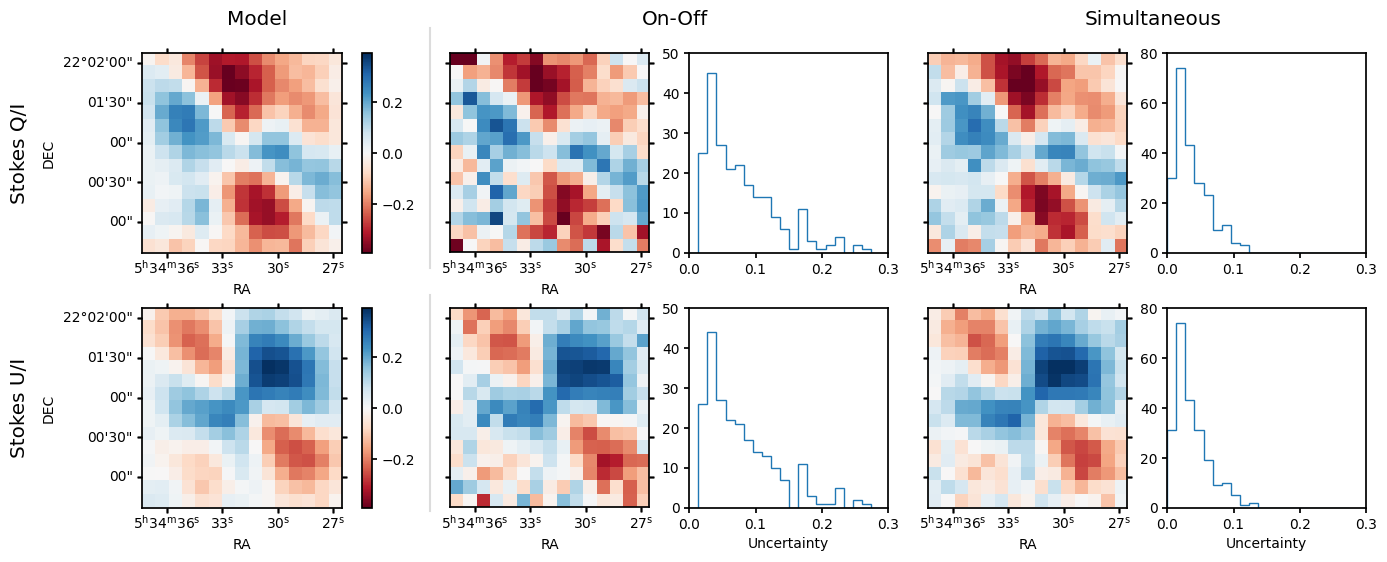}    \caption{On-off and simultaneous fits for nebula polarization from a simulated 100ks Crab observation. Left column is the model maps of Stokes $q = Q/I$ and $u = U/I$. Their color bar is common to all panels. Beside each reconstructed map (middle and right columns) is a histogram of the uncertainties of the polarization values. Note that simultaneous fitting produces maps that more cleanly resemble the model with smaller uncertainties.}
    \label{fig:nebula_sim}
\end{figure*}

Figures \ref{fig:pulsar_sim} and \ref{fig:nebula_sim} show the reconstructed polarization of the two fitting methods as well as the input model. To compare the two methods, we used three summary statistics elaborated below: the median error bar size, the GoF, and the number of measurements. For the pulsar measurements, the median error bars decrease by $\times 1.20$, averaged between q and u; there are also $(29/23=1.26)\times$ more measurements in the simultaneous case. We can further characterize the systematic errors in the recovery by a `Goodness of Fit' 

$$\mathrm{GoF}_q=  \{ \sum_n [( q-q_{\rm mod})/\sigma_q]^2 / n 
 \} ^{1/2}$$ 

\noindent and similarly for $u$, with an average $\times 1.08$ improvement in recovery of the original model, even relative to the improved errors. Thus, in total, simultaneous fitting can be considered to provide a $1.20\times 1.26 \times 1.08 \approx 1.63\times$ improvement in recovery of the pulsar polarization.

For the nebula, since simultaneous fitting uses nearly all the data away from the pulsar, especially off the brightest portion of the peaks, the effective nebula exposure is larger by $1/\Delta\phi_{\rm off} \approx 4.9\times$ than in the simple ``off" portion of the pulse phase used to map the nebula in the on-off method. Further, the method takes account of the small expected pulsar flux in off phases to provide a cleaner measurement of the true nebulae structure very close to the pulsar. The polarization maps in Figure \ref{fig:nebula_sim} show substantially improved source recovery and decreased typical uncertainty in the recovered pixels. Quantitatively, the median error bars decrease by $\sim 2\times$ for both q and u. Although the GoF for simultaneous fitted nebula is slightly larger than the on-off nebula, since the uncertainty is halved, this actually means that simultaneous fitting is able to recover polarization values closer to truth.

\section{Application to First IXPE Crab Observation}
\label{sec:data}

In February and March 2022, IXPE made a $\sim$91 ks observation of the Crab, its first PWN source. At the time, the mirrors were misaligned from the pointing axis by $\sim$3 arcmin, and we do not have energy-dependent response functions calibrated for this offset. This (and the incursion of Poisson statistics in low count bins, see below) led us to analyze in a single 2--8 keV {\it IXPE} energy bin.

Moreover, it has been discovered that errors in the present track reconstruction of {\it IXPE} photon conversion points are correlated with the initial direction of the photoelectron track, and hence, with the inferred event polarization. This leads to ``polarization leakage," which induces polarized fringes about sharply localized X-ray sources (point source and compact nebulae). As described by \citet{2023arXiv230200346B}, these fringes average away for point sources, and hence do not affect aperture polarization measurements, but do affect the edges of extended sources, such as the Crab Nebula. The paper also gives a prescription for correcting these fringes, assuming a smooth Gaussian blurring of the point sources.

We implement here an improved version of this correction, using more detailed {\it IXPE} PSFs for each mirror assembly, derived from ground calibration data and on-sky observations of point sources. The mirror PSF are lightly smoothed to suppress numerical noise. We then compute maps of the Hessian terms $H_{xx}$, $H_{yy}$ and $H_{xy}$. On-sky images have residual blur beyond the mirror PSFs, produced by incomplete aspect correction and imperfect estimation of the photon conversion points. We treat these as simple Gaussian blurs with $\sigma_{\rm G}=2.1^{\prime\prime}, \,  1.4^{\prime\prime}, \,  1.2^{\prime\prime}$ for detector units (DUs) 1, 2, 3 respectively. In addition, the correlation between the conversion point and the EVPA induces a `leakage' blur, which is energy dependent and grows at large photon energies. We model the effect as $\sigma_{\rm L} = (10+3\Delta E)^{1/2}$ arcsec, with $\Delta E =E_{\rm keV}-4$keV for photon energy above $4$\,keV and $\Delta E=0$ below. This is common to all detectors. The effective PSFs for unpolarized sources are 

\begin{equation}
\begin{split}
    P_I^*(\vec{\mathbf{x}}) &= (P_M \star G(\sigma_G ))(\vec{\mathbf{x}}) \\
        & \qquad + \frac{\sigma _L^2}{4} (H_{xx}(\vec{\mathbf{x}}) + H_{yy}(\vec{\mathbf{x}}))\\
    P_Q^*(\vec{\mathbf{x}}) &= \frac{\sigma _L^2}{4} (H_{xx}(\vec{\mathbf{x}}) - H_{yy}(\vec{\mathbf{x}}))\\
    P_U^*(\vec{\mathbf{x}}) &= \frac{\sigma _L^2}{2} (H_{xy}(\vec{\mathbf{x}}))
\end{split}
\end{equation}

\noindent with $P_M$ the appropriate mirror PSF and $\star\ G$ the symmetric Gaussian convolution. For a polarized source, $P_{I,Q,U}$ are mixed by the blurring effect as

\begin{equation}
\begin{split}
    P_I(\vec{\mathbf{x}}) &= P_I^* (\vec{\mathbf{x}}) + {1 \over 2}[q_{\rm src} P_Q^*(\vec{\mathbf{x}}) + u_{\rm src} P_U^*(\vec{\mathbf{x}})] \\
    P_Q(\vec{\mathbf{x}}) &= q_{\rm src} P_I^*(\vec{\mathbf{x}}) + P_Q^*(\vec{\mathbf{x}}) \\
    P_U(\vec{\mathbf{x}}) &= u_{\rm src} P_I^*(\vec{\mathbf{x}}) + P_U^*(\vec{\mathbf{x}}),
\end{split}
\end{equation}

\noindent where the $q_{\rm src}$ and $u_{\rm src}$ are the estimates of the true source polarization.

In our implementation, we iterate, starting by fitting to raw {\it{IXPE}} $I$, $Q$, $U$ data, to arrive at leakage-corrected polarization fits in $\sim$5 steps. In the method, we form leakage-corrected $q$ and $u$ maps for each phase and energy bin and detector. These are fed to the simultaneous fitting minimizer, which separates the nebula and pulsar signals. In our case, during fitting, the energy bins are collapsed. The results are final derived nebula $q$ and $u$ maps and pulsar phased $q$ and $u$ values, corrected for the spatially-dependent polarization leakage. In simulation, our prescription has shown to improve recovery of the original $q_{src}$ and $u_{src}$. For details, see \cite{Dinsmore:2023}. Our analysis below includes this correction, which can make substantial modifications to the nebula polarization map near the pulsar and at the outer edges. Near-pulsar corrections, in turn, feed back to changes in the phase-resolved polarization.

Using the same procedure as for simulated data, we performed on-off and simultaneous fits to the Crab observation. For on-off, we used the same phase bins, area, and off-pulse range as the initial IXPE discovery paper on the Crab (\citealt{2023NatAs.tmp...74B}) and replicated those results.

For simultaneous fitting, we used a 150$^{\prime\prime}$ $\times$ 150$^{\prime\prime}$ area, binned into a 15 $\times$ 15 pixel grid. Using the phase bins of \citet{2023NatAs.tmp...74B}, we also obtain significant polarization detection in the main peak (P1) phase bin $[0.12,0.14]$ with a higher significance of PD = $15.1 \pm 2.1\%$. The second peak (P2) phase bin $[0.515, 0.545]$ was also found to be significantly polarized with PD = $8.8 \pm 2.8\%$; these errors decrease from the on-off values.

\begin{figure}
    \centering
    \includegraphics[width=\linewidth]{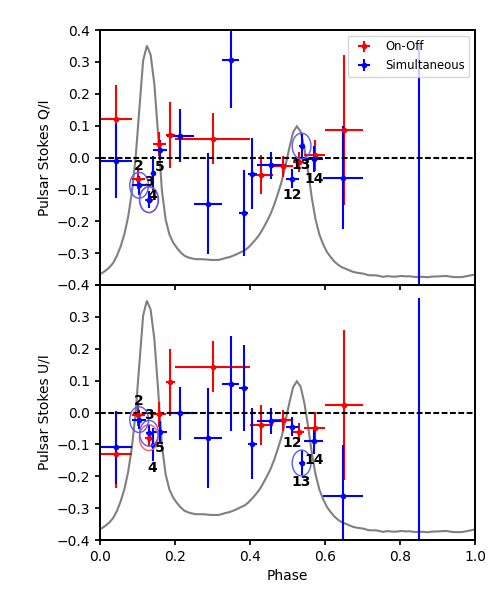}
    \caption{{\it IXPE} Crab pulsar Stokes Q/I and U/I, after leakage correction, from on-off and simultaneous fitting. Statistically significant detections ($> 3\sigma$) are circled red and blue, respectively. With simultaneous fitting, due to the extra signal we extract from the data, we can afford to use finer (and more) phase bins, which helps us capture the polarization sweep. Both methods have $> 5\sigma$ detection at the first peak. The X-ray light curve is displayed in gray.}
    \label{fig:pulsar_data}
\end{figure}

\begin{figure}
    \centering
    \hspace{-5mm}
    \includegraphics[width=\linewidth]{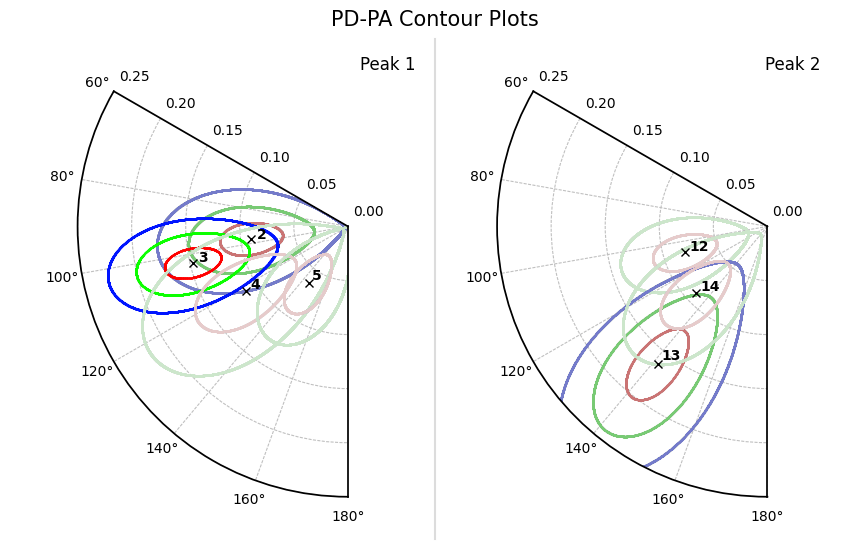}
    \caption{{\it IXPE} Crab pulsar PD-PA contour plots showing $1\sigma, 2\sigma, 3\sigma$ (red, green, blue) contours for select simultaneous-fitted bins around the two peaks. Phases with greater than $5\sigma$ detection are the brightest in color and fade for the $>2\sigma$ bins. Numbers correspond with the phase bins numbered in Figure \ref{fig:pulsar_data}. For P1, a CCW sweep can be inferred; P2 requires more bins to show a polarization pattern.}
    \label{fig:pulsar_data_contours}
\end{figure}

Seeing that we can measure polarization with higher significance, we decided to use smaller bins around the peak phases, where we can expect fast sweeps in polarization angle, to better measure the polarization curve.
Here we employ 16 phase bins, as compared to the 11(+off) phase bins used in the on-off analysis. With these new bins and corrections, we obtain refined Stokes \textit{q} and \textit{u} phase points (Figures \ref{fig:pulsar_data}, \ref{fig:pulsar_data_contours} and Table \ref{tab:simul_pol}), which may be compared with the standard on-off results. 

Polarization angle (PA) and degree (PD) are useful for comparison with models, but these do not have simple Gaussian errors. We can however plot the confidence regions for the phase bins around the two peaks (Figure \ref{fig:pulsar_data_contours}), where three phase bins have above $3\sigma$ significance, and four more are above $2\sigma$. In P1, we see a counterclockwise (CCW) sweep from $\sim 100^{\circ} - 150^{\circ}$. P2 hints at a CCW sweep as well, though since it only has one significant bin, more data is required to see this.

\begin{figure}[h!!]
    \centering
    \includegraphics[width=\linewidth]{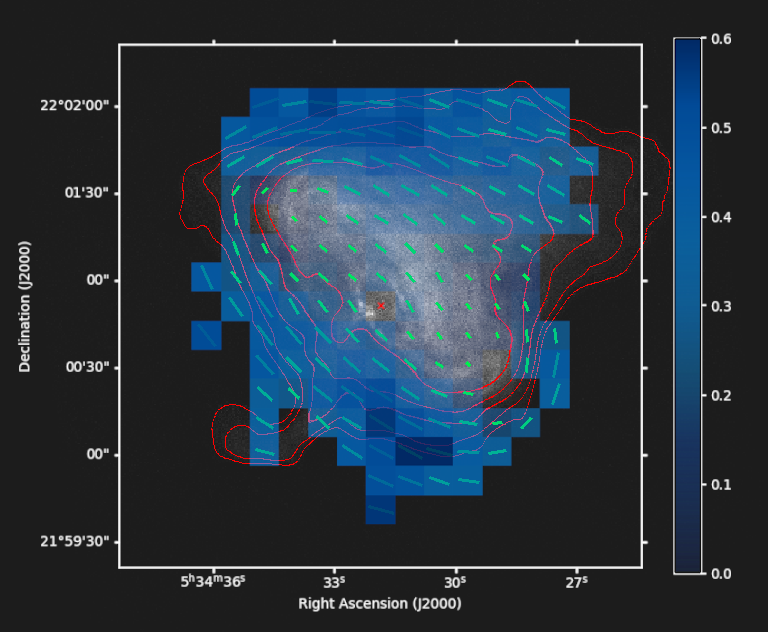}
    \caption{{\it IXPE} Crab nebula polarization map, after leakage correction. Green lines indicate magnetic field direction (perpendicular to the electric field PA) and the tile color indicates PD. Contour intensity levels (red) derived from the contemporaneous Chandra Crab observation (OBSID 23539), both displayed in the background, are included for reference; the pulsar position is marked with a red cross.}
    \label{fig:nebula_data}
\end{figure}

Figure \ref{fig:nebula_data} shows the reconstructed nebula PD map, cut at 4.7$\sigma$ significance, with green bars showing PA and magnetic field direction. As seen in \citealt{2023NatAs.tmp...74B}, the PD is reduced at the sides of the torus where the PA sweeps rapidly through 180$^\circ$. With leakage correction, these feature are enhanced, appearing as PD holes on opposite sides of the nebula. In general, the PD is largest towards the northern and southern edges of the nebula; in part, this may be because such regions have toroidal PA at a similar angle across the PSF, minimizing polarization beam dilution.  A highly polarized PD $>$ 50\% region west of the jet is also present. Generally, the magnetic field lines follow the filamentary structure partly visible in the background image.

\begin{table*}
\begin{center}
\hspace*{-3.7cm}
\begin{tabular}{ | c | r | r | r| r | r || c | r | r | r | r | r |} 
 \hline
 \multicolumn{6}{|c||}{On-Off} & \multicolumn{6}{c|}{Simultaneous} \\
 \hline
 & Phase & q & u & PD (\%) & PA ($^\circ$) & & Phase & q & u & PD (\%) & PA ($^\circ$)\\
 \hline
 \multirow{4}{*}{P1} & \multirow{4}{1.0cm}{$0.120 - 0.140$} & \multirow{4}{1.2cm}{$-0.132 \pm 0.025$} & \multirow{4}{1.2cm}{$-0.079 \pm 0.025$} & \multirow{4}{1.0cm}{$15.4 \pm 2.5$} & \multirow{4}{1.0cm}{$105 \pm 18$} & 2 & $0.085-0.124$ & $-0.087 \pm 0.029$ & $-0.022 \pm 0.029$ & $9 \pm 3$ & $97 \pm 9$ \\
 \cline{7-12}
 & \multirow{3}{1.0cm}{} & \multirow{3}{1.2cm}{} & \multirow{3}{1.2cm}{} & \multirow{3}{1.0cm}{} & \multirow{3}{1.0cm}{} & 3 & $0.124-0.138$ & $-0.132 \pm 0.027$ & $-0.065 \pm 0.027$ & $15 \pm 3$ & $103 \pm 5$ \\
 \cline{7-12}
 & \multirow{2}{1.0cm}{} & \multirow{2}{1.2cm}{} & \multirow{2}{1.2cm}{} & \multirow{2}{1.0cm}{} & \multirow{2}{1.0cm}{} & 4$^{*}$ & $0.138-0.142$  & $-0.048 \pm 0.053$ & $-0.100 \pm 0.053$ & $11 \pm 5$ & $122 \pm 14$ \\
 \cline{7-12}
 & \multirow{1}{1.0cm}{} & \multirow{1}{1.2cm}{} & \multirow{1}{1.2cm}{} & \multirow{1}{1.0cm}{} & \multirow{1}{1.0cm}{} & 5$^{*}$ & $0.142-0.179$  & $ 0.028 \pm 0.032$ & $-0.059 \pm 0.032$ & $6 \pm 3$ & $145 \pm 14$\\
 \hline
 \hline
 \multirow{3}{*}{P2} & \multirow{3}{1.0cm}{$0.515 - 0.545$} & \multirow{3}{1.2cm}{$-0.013 \pm 0.030$} & \multirow{3}{1.2cm}{$-0.06 \pm 0.030$} & \multirow{3}{1.0cm}{$6.3 \pm 3.0$} & \multirow{3}{1.0cm}{$129 \pm 14$} & 12$^{*}$ & $0.495-0.530$ & $-0.066 \pm 0.031$ & $-0.044 \pm 0.031$ & $8 \pm 3$ & $107 \pm 11$ \\ 
 \cline{7-12}
 & \multirow{2}{1.0cm}{} & \multirow{2}{1.2cm}{} & \multirow{2}{1.2cm}{} & \multirow{2}{1.0cm}{} & \multirow{2}{1.0cm}{} & 13 & $0.530-0.545$ & $0.036 \pm 0.039$ & $-0.158 \pm 0.039$ & $16 \pm 4$ & $141 \pm 7$ \\
 \cline{7-12}
 & \multirow{1}{1.0cm}{} & \multirow{1}{1.2cm}{} & \multirow{1}{1.2cm}{} & \multirow{1}{1.0cm}{} & \multirow{1}{1.0cm}{} & 14$^{*}$ & $0.545-0.595$ & $-0.006 \pm 0.040$ & $-0.090 \pm 0.040$ & $9 \pm 4$ & $133 \pm 13$ \\
 \hline
\end{tabular}
\end{center}
\caption{\label{tab:simul_pol} Numerical comparison of published on-off measurements \citep{2023NatAs.tmp...74B} and simultaneous fit measurement of pulsar peak polarizations. $^{*}$Note: These low significance bins have substantial PD-PA covariance; see Figure \ref{fig:pulsar_data_contours} for 2D error contours.}
\end{table*}

\section{Discussion}
\label{sec:dis}

In this paper, we show that simultaneous fitting has several merits. As tested with simulated data, it provides improved recovery of the pulsar polarization sweep with smaller statistical uncertainties on both pulsar and nebula measurements. This allows us to use smaller phase bins to more finely resolve the pulsar polarization. Our method essentially uses externally measured, high precision data on the time and spatial intensity structure to provide a weighting that improves extraction of the polarization signal. Perhaps the greatest limitation in the present implementation is our assumption of Gaussian statistics. While this allows an efficient linear algebra solution for the many pulse phase and nebula image polarization values, it does break down when the counts in  a given bin are too low. We have noticed that such Poisson effects may be important in the low count outskirts of the PWN, although in the computations presented here $<2.5\%$ of the $16\times 15\times 15 \times 3= 10,800$ data bins have $<10$ counts. However, this does limit our ability to extend our decomposition to be fully energy-dependent, as the higher energy {\it IXPE} bins often have low count rate. With additional Crab exposure it will be straight-forward to extend the analysis to a modest number of spectral bins. 

\begin{figure}
    \hspace{-6mm}
    \includegraphics[width=1.1\linewidth]{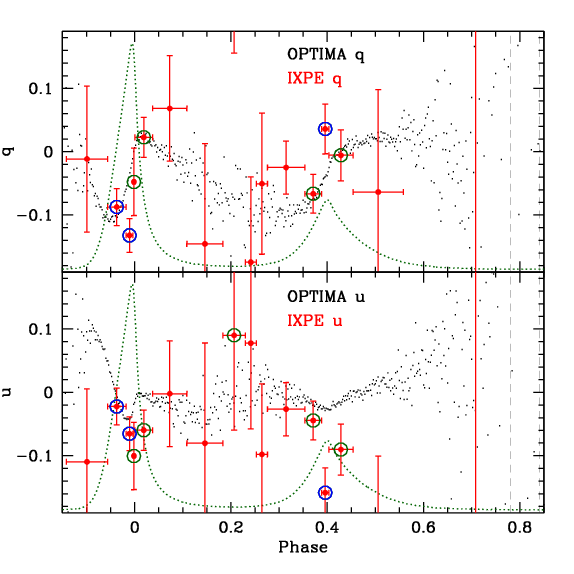}
    \vskip -5mm
    \caption{{\it IXPE} X-ray polarization measurements, after leakage correction, compared with optical values.
    Optical $q=Q/I$ and $u=U/I$ from off-pulse subtracted OPTIMA observations are shown as black dots (with the optical light curve in green for phase reference). The average OPTIMA $q$ and $u$ fluxes will be zero in the dashed line-delimited off-pulse region. Red error bars show the {\it IXPE} measurement uncertainties. Phase bins whose PD exceeds $3\sigma$ significance are circled in blue, those exceeding $2\sigma$ in dark green. Note that while optical and X-ray $q$ values at the second peak are consistent, there are substantial differences for the other peak polarizations.} 

    \label{fig:pulsar_data_opt}
\end{figure}

The polarization of the optical Crab pulsed emission has been measured repeatedly in the 50 years since its discovery. The measurements of \citet{Slowikowska:2009} with the OPTIMA fiber photometer are of particularly high quality. The central fiber used for these measurements had a diameter of $2.3^{\prime\prime}$ on the sky and thus included both the Crab pulsar and nebular emission such as the ``inner knot". Accordingly these authors define a ``background" phase of $\phi=0.78-0.84$ (dashed lines in Figure \ref{fig:pulsar_data_opt}) and subtract the Q and U fluxes from this phase to get the pulsed Crab emission. This turns the optical Crab curves into ``on-off" measurements. In Figure \ref{fig:pulsar_data_opt}, we compare this pulsed optical polarization signal (measurements kindly supplied by G. Kanbach) with our new {\it IXPE} measurements. Alignment of the optical and X-ray curves to the radio phase convention was checked via their light curves, with the X-ray peak at $\phi=0.99$ and the optical peak at $\phi=0.994$ \citep{Enoto:2021}.

There have been several attempts to compare the OPTIMA polarization data with theoretical pulsar models \citep[e.g.][]{2004ApJ...606.1125D}, none particularly satisfactory. Clearly some ingredients are missing in our basic understanding the pulse emission. In particular, absorption effects or contributions from multiple emission regions may introduce complications beyond the essentially geometric models that have been applied to date. Such effects should differ between the X-ray and optical bands. So it is encouraging that we do see statistically significant differences between the optical OPTIMA and {\it IXPE} polarization curves. 
Most notably, the P1 $q$ sweep starts from larger negative values and is delayed in phase. Negative $u$ values also persist later in P1. Similarly $u$ is much more negative in the core of P2. We do not find any significant measurements in the pulse minimum bin which is substantially nebula dominated, with a maximum of $\sim 20\%$ pulsar counts in the central pixel (and steeply tapering in adjacent pixels). Additional signal-to-noise (S/N) can allow better separation of the pulse signal near this phase. Near the peaks the polarization values have significant sensitivity to our choice of bin boundaries, so we suspect that unresolved rapid sweeps still suppress the polarization. With additional exposure, {\it IXPE} can measure several more phase bins with good precision. Comparison of the optical and X-ray signals can then give new insights into the geometry of the pulse emission zones. By September 2023, {\it IXPE} will conclude a follow-up 300 ks observation of the Crab, which, combined with the current dataset and event quality weighting, should more than double the S/N presented here.

\section{Conclusion}
\label{sec:concl}
Analysis of the first {\it IXPE} Crab observation using simultaneous fitting has improved our measurements of the Crab polarization. Compared to the original paper (\citealt{2023NatAs.tmp...74B}), we recover more bins of significant pulsar polarization and are able to use a finer phase resolution to see departures from the well-measured optical polarization. Nebula features, such as the PD holes at the edge of the torus, are better recovered. With the additional S/N from planned Crab follow-up exposure, we can substantially extend these gains. The method is of course general and can be applied to any phase varying source embedded in extended emission. We anticipate that application to other {\it IXPE} sources, such as MSH 15-5(2) and possibly B0540 and Vela, can provide improved measurements as well.

\vspace{-0.5cm}
\acknowledgments

This work was supported by NASA under grant NNM17AA26C.

The Imaging X-ray Polarimetry Explorer (IXPE) is a joint US and Italian mission. The US contribution is supported by the National Aeronautics and Space Administration (NASA) and led and managed by its Marshall Space Flight Center (MSFC), with industry partner Ball Aerospace (contract NNM15AA18C). The Italian contribution is supported by the Italian Space Agency (Agenzia Spaziale Italiana, ASI) through contract ASI-OHBI-2017-12-I.0, agreements ASI-INAF-2017-12-H0 and ASI-INFN-2017.13-H0, and its Space Science Data Center (SSDC) with agreements ASI-INAF-2022-14-HH.0 and ASI-INFN 2021-43-HH.0, and by the Istituto Nazionale di Astrofisica (INAF) and the Istituto Nazionale di Fisica Nucleare (INFN) in Italy. 

This research used data products provided by the IXPE Team (MSFC, SSDC, INAF, and INFN) and distributed with additional software tools by the High-Energy Astrophysics Science Archive Research Center (HEASARC), at NASA Goddard Space Flight Center (GSFC).

\bibliography{references}{}
\bibliographystyle{aasjournal}

\end{document}